\documentstyle[namedreferences,proceedings,psfig]{crckapb}

\ifx \undefined\vctr
  \def\vctr#1{\ifmmode\mathchoice{\mbox{\boldmath$\displaystyle#1$}}
  {\mbox{\boldmath$\textstyle#1$}}
  {\mbox{\boldmath$\scriptstyle#1$}}
  {\mbox{\boldmath$\scriptscriptstyle#1$}}\else
  \hbox{\boldmath$\textstyle#1$}\fi}
\fi

\newcommand{\Rs}{r_{\rm s}}

\newcommand{\CITEX}[1]{\citeauthor{#1} (\citeyear{#1})}

\begin{opening}
\title{The microlensing events in Q2237+0305A:\protect\\
  No case against small masses/large sources}

\author{Stein Vidar Hagfors Haugan}
\institute{Institute of Theoretical Astrophysics, University of Oslo\\
  Pb. 1029, Blindern\\
  N-0315 OSLO\\
  {\tt http://www.uio.no/\~\/steinhh/index.html}}

\end{opening}

\runningtitle{The microlensing events in Q2237+0305A}
\begin{document}


\section{Introduction}
\CITEX{Witt-Mao94} claim that the data reported by \CITEX{Racine92}
contains ``a quite well sampled M-shaped double event in image A of
0.3 and 0.4 mag, respectively''. They further state that the very low
average mass scenario put forward by \CITEX{Refsdal-Stabell91} does
not predict ``well-resolved {\em asymmetric} events, as have been
observed in image~A''.

The first peak has only six sampling points, with all but one point
clustered on one side of the peak. The second peak has 5 sampling
points, with {\em all} points clustered on one side of the peak. The
degree of asymmetry is thus very hard to quantify.

\section{The Large Source Model}
\label{sec:lowmass-model}

I have studied a large number of large source lightcurves
\cite{Haugan94} produced with the rayshooting method.  A Gaussian
source profile $I(\vctr{y}) \propto \exp(-|\vctr{y}|^2/\Rs^2)$ was
used, where $\vctr{y}$ is the dimensionless position relative to the
source center, and $\Rs$ is the dimensionless source size.

Large source models do not typically produce asymmetric events, but
they certainly do occur. In order to highlight the problems of using
isolated events in order to determine the normalized source size,
Fig.~1 shows a curve similar to the one appearing in \CITEX{Racine92}
superposed on simulated lightcurves with large sources.  The
lightcurve parameters $\kappa_*$, $\gamma$, and $\Rs$, are indicated.
Positive $\gamma$ indicates that the large, elongated caustic
structures are oriented along the source track. A time-reversed
version of the spline curve is also supplied to aid the eye.

\begin{figure}[htbp]
  \vspace{-.3cm}
  \centerline{\psfig{figure=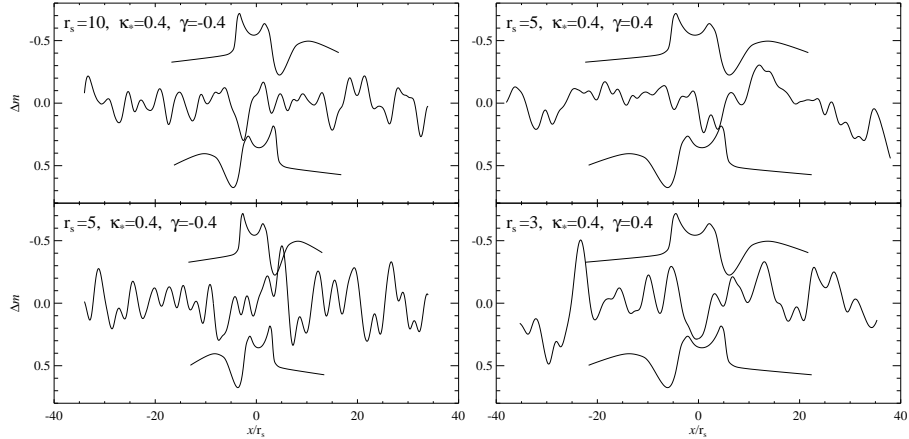,width=\textwidth}}
  \vspace{-.4cm}
  \caption{Lightcurves from simulations with a curve similar to
    the one in \protect\CITEX{Racine92} superimposed. Abscissa values
    are in units of the source size $\Rs$.}
  \label{fig:lightcurves}
\end{figure}

Although searching for exact replicas of the observed lightcurve among
simulated lightcurves is useless, a comparison by eye can easily be
done.
 Although lightcurves
with large sources lack the clear M-shaped events of lightcurves with
small sources ($\Rs \ll 1$), the peaks may very well be asymmetric to
the extent indicated by the observations.

\section{Conclusion}
\label{sec:conclusion}

Based on the above arguments, the exclusion of models with a large
source or low average masses is {\em not} justified from the 1988-90
events.  Further observations and analysis should therefore not be
concentrated solely on interpreting the lightcurves from the
perspective of (very) small sources.

\section*{Acknowledgments}
The author wishes to thank Rolf Stabell, Sjur Refsdal and Per Barth
Lilje for helpful comments on the manuscript.


\begin{thebibliography}{}

\bibitem[\protect\citeauthoryear{Haugan}{1994}]{Haugan94}
Haugan, S. V.~H. 1994,
\newblock {\em Master's thesis}, University of Oslo, Institute of Theoretical
  Astrophysics, Pb 1029 Blindern, N-0315 Oslo, Norway

\bibitem[\protect\citeauthoryear{Racine}{1992}]{Racine92}
Racine, R. 1992, ApJ, 395, L65

\bibitem[\protect\citeauthoryear{Refsdal \& Stabell}{1991}]{Refsdal-Stabell91}
Refsdal, S., Stabell, R. 1991, A\&A, 250, 62

\bibitem[\protect\citeauthoryear{Witt \& Mao}{1994}]{Witt-Mao94}
Witt, H.~J., Mao, S. 1994, ApJ, 429, 66

\end{thebibliography}
\end{document}